\documentclass[12pt,twoside,epsfig,here,graphicx]{article}
\usepackage{geometry}
\usepackage{epsfig}
\usepackage{graphics}
\usepackage{extramarks}

\begin{document}                                        
\begin{center}                                          
\large{\bf \boldmath Interaction of $K^-$-mesons with light nuclei} \\
\vspace{3ex}                                            
\underline{L.A. Kondratyuk}$^{1}$\\
           V.Yu. Grishina$^{2}$\\
           M. B\" uscher$^{3}$\\
\vspace{2ex}                                            
\small{
$^{1}$ Institute of Theoretical and Experimental Physics, B.\
Cheremushkinskaya 25, 117218 Moscow, Russia \\
$^{2}$ 60th October Anniversary Prospect 7A, 117312 Moscow, Russia   \\
$^{3}$ Institut f\"ur Kernphysik, Forschungszentrum J\"ulich,
D-52425 J\"ulich, Germany           
}
\end{center}                                            
\vspace{1ex}                                            
\normalsize{                                            

\section*{Introduction}

Low-energy ${\bar K}N$ and ${\bar K}A$ interactions have gained
substantial interest during the last two decades. Data on the $K^-p$
scattering length $a(K^-p)$ from KEK~\cite{Ito,Iwasaki}
\begin{equation}
   a(K^-p)=(-0.78 \pm 0.18) + i(0.49 \pm 0.37) ~{\mbox{fm}},
   \label{Iwasaki}
\end{equation}
and the DEAR experiment at Frascati~\cite{Guaraldo1}
\begin{eqnarray}
  & a(K^-p)=&(-0.468 \pm 0.090\,({\mbox{stat.}}) \pm
  0.015\,({\mbox{syst.}})) \nonumber \\ & &+i (0.302 \pm
  0.135\,({\mbox{stat.}}) \pm
  0.036\,({\mbox{syst.}}))~{\mbox{fm}}
\label{Guaraldo1}
\end{eqnarray}
show that the energy shift of the 1$s$ level of kaonic hydrogen is
repulsive, Re\,$a(K^-p)<0$. Nevertheless, it is possible that the
actual $K^-p$ interaction is attractive if the isoscalar
$\Lambda(1405)$ resonance is a bound state of the ${\bar K}N$
system~\cite{Dalitz,Weise1}. A fundamental reason for such a scenario
is provided by the leading order term in the chiral expansion for the
$K^-N$ amplitude which is attractive.

Furthermore, very recently a strange tribaryon $S^0(3115)$ was
detected in the interaction of stopped $K^-$-mesons with
$^4$He~\cite{Suzuki}. The width of this state was found to be less
than 21~MeV. According to Ref.~\cite{Dote} this state can be
interpreted as a candidate of a deeply bound state $({\bar
K}NNN)^{Z{=}0}$ with $I{=}1,I_3{=}{-}1$. It is clear that further
searches for bound kaonic nuclear states as well as new data on
the interactions of $\bar K$-mesons with lightest nuclei are thus
of great importance.


\section*{The \boldmath $K^- d$, $K^-\,^3{\mathbf{He}}$, and $K^- \alpha$
scattering lengths and  loosely bound  $K^-$-nucleus states}

Calculations of the $K^- d$ scattering length have recently been
performed within the Multiple-Scattering Approach
\cite{Grishina_a0_04,Deloff00,Kamalov01} as well as with Faddeev
Equations \cite{Deloff00,Bahaoui03}. The results of
Refs.~\cite{Grishina_a0_04} and \cite{Deloff00} if using the same
input are in good agreement. Moreover, the calculations of Barrett and
Deloff~\cite{Deloff00} within the framework of Faddeev equations gave
practically the same result as in the Multiple Scattering Approach,
with the value for $A(K^- d)$ in the range $-(0.75 \div 0.85) + i
(1.10 \div 1.23)$~fm.

In Ref.~\cite{Sibirtsev04} the real and imaginary parts of the ${\bar
K^0}d$ scattering length have been extracted from the data on the
${\bar K^0}d$ mass spectrum obtained from the reaction $pp{\to}d{\bar
K^0}K^+$ measured recently at COSY~\cite{Kleber03}.  Upper limits on
the $K^-d$ scattering length have been found, namely $\mathrm{Im} \,
A(K^- d){\le}$1.3~fm and $|\mathrm{Re}\, A(K^- d)|{\le}$1.3~fm. It
has also been shown that the limit for the imaginary part of the
$K^-d$ scattering length is strongly supported by data on the total
$K^-d$ cross sections. The results for the imaginary part of the
$K^-d$ scattering length from Refs.~\cite{Kamalov01} and
\cite{Bahaoui03} violate the upper limit found in Ref.~\cite{Sibirtsev04}.

The calculations of the $K^-\alpha$ and $K^-\, ^3{\mathrm{He}}$
scattering lengths have been performed using five parameter sets for
the ${\bar K}N$ lengths shown in Table~\ref{Tab2}. The results from a
$K$-matrix fit (Set 1), separable fit (Set 2) and the constant
scattering length fit (CSL) denoted as Set 3 were taken from
Ref.~\cite{Barret}. We also study the CSL fit from
Conboy~\cite{Conboy} (Set 4). Recent predictions for ${\bar K}N$
scattering lengths based on the chiral unitary approach of
Ref.\cite{Oller} are denoted as Set~5.

\begin{table*}[htb]
\begin{center}
\begin{tabular}{|l|c|l|l|l|l|}
\hline  Set & Ref. & $\vphantom{\displaystyle \frac
{A^{I}}{B^{I}}}a_0(\bar{K}N) [{\mbox{fm}}]$&
$a_1(\bar{K}N)[{\mbox{fm}}]$& $A(K^- \alpha) [{\mbox{fm}}]$ &
$A(K^-\,^3{\mathrm{He}}) [{\mbox{fm}}]$
\\ \hline
1 & \cite{Barret} & $-1.59+i0.76$ & $0.26 + i0.57$ & $-1.80+ i0.90$
& $-1.50+i0.83$ \\ \hline
2 & \cite{Barret} & $-1.61+i0.75$ & $0.32 + i0.70$ & $-1.87 + i 0.95$
& $-1.55+i0.90$ \\ \hline
3 & \cite{Barret} & $-1.57+i0.78$ & $0.32 + i0.75$ & $-1.90 + i 0.98$
& $-1.58 + i0.94$ \\ \hline
4 & \cite{Conboy} & $-1.03+i0.95$ & $0.94 + i0.72$ & $-2.24+ i 1.58$
& $-1.52 + i1.80$ \\ \hline
5 & \cite{Oller} & $-1.31+i1.24$ & $0.26 + i0.66$ & $-1.98+ i 1.08$
& $-1.66 + i1.10$ \\ \hline
\end{tabular}
\end{center}
\caption{\label{Tab2} The $K^- \alpha$ scattering length from
Ref.~\cite{Grishina05} and new results for $K^-
\,{}^{3}\mathrm{He}$, calculated for different
sets of the elementary $\bar K N $ amplitudes $a({\bar{K}N})$
($I=0,1$).}
\end{table*}

The results of the calculations are listed in the last two columns
of Table~\ref{Tab2}. These results are very similar for Sets 1--3.
The $K^- \alpha$ and $K^- \, ^3{\mathrm{He}}$ scattering lengths
are in the range $A(K^-\alpha)=-(1.8 \div 1.9)+i(0.9 \div
0.98)$~fm and $A(K^- \,^3{\mathrm{He}})=-(1.5 \div 1.58)+i(0.83
\div 0.94)$~fm, respectively. The results for Set~4 are quite
different: $A(K^-\alpha)=-2.24+i1.58$~fm and $A(K^- \,
{}^3\mathrm{He})= -1.52+i1.80$~fm. The calculations with Set~5 are
close to the results obtained with Sets 1--3.

Unitarizing the constant scattering length,  we can reconstruct
the $K^-X$ scattering amplitude within the zero range
approximation (ZRA) as
\begin{equation}
 f_{K^- X}(k)=
\left[A(K^-X)^{-1} - ik\right]^{-1}, \label{f_KHepole}
\end{equation}
where $X=\alpha$ or ${}^3\mathrm{He}$, $k{=}k_{K^-X} $ is the
relative momentum of the $K^-X$ system. The denominator of the
amplitude of Eq.(\ref{f_KHepole}) has a zero at the complex energy
$E^*=E_R - i\Gamma_R/2= k^2/(2\mu)$,
where $E_R$ and $\Gamma_R$ are the binding energy and width of a
possible $K^-X$ resonance, respectively. Here $\mu$ is the reduced
mass of the $K^-X$ system.

In case of the $K^- \alpha$ system we find for Sets 1 and 4 a pole at
the complex energies $E^*{=}(-6.7{-}i 18/2)$ MeV and $E^*{=}(-2.0{-}i
11.3/2)$ MeV, respectively. The calculations with Set 5 also result in
a loosely bound state, $E^*{=}(-4.8{-}i 14.9/2)$ MeV. Similar results
have been obtained for the $K^-\,^{3}{\mathrm{He}}$ system. Note that
assuming a strongly attractive phenomenological $\bar K N$ potential,
Akaishi and Yamazaki~\cite{Akaishi} predicted a deeply bound ${\bar
K}\alpha$ state at $E^*{=}(-86{-}i 34/2)$ MeV, which is far from our
solutions. This problem can be resolved assuming that the loosely and
deeply bound states are different eigenvalues of the ${\bar K}\alpha$
effective Hamiltonian. Our model for the ${\bar K}\alpha$ scattering
amplitude is valid only near threshold, {\it i.e.\/} when
$kA(\bar{K}\alpha){\ll}1$. The ZRA can not be applied for the
description of deeply bound states when the pole of the scattering
amplitude is located far away from the threshold. If the same
procedure were applied to the $K^-\,^3$H system we would find a
similar loosely bound state. This state together with recently
discovered deeply bound state, the $S^0$(3115), can be considered as
different eigenvalues of the $K^-\,^3$H effective Hamiltonian. In any
case it is very important to measure the $s$-wave ${\bar K}\alpha$
scattering length in order to clarify the situation concerning the
existence of bound ${\bar K}\alpha$ states.

\section*{The {\boldmath $K^-\alpha$} FSI in the reaction {\boldmath $dd{\to}\alpha{K^- K^+}$}}
In Refs.~\cite{Grishina05,Buescher02} it was argued that the
reaction $dd \to \alpha K^- K^+ \label{ddHeKK}$ near threshold is
sensitive to the $K^-\alpha $ final state interaction. We
calculated the ~$K^-\alpha$ invariant mass spectrum at excess
energy $50$~MeV. The result is shown in Fig.~\ref{fig1}. The solid
line shows the calculations for pure phase space, {\it i.e.\/} for
a constant production amplitude and neglecting FSI effects. The
dash-dotted and dashed lines show the results obtained for the
$K^-\alpha$ FSI calculated with the parameters of Sets 1 and 4,
respectively. All lines are normalized to the same total cross
section of 1~nb. It is clear that the FSI significantly changes
the $K^-\alpha$ mass spectrum.

\begin{figure}[htb]
\vspace*{-2mm}
\centerline{\epsfig{file=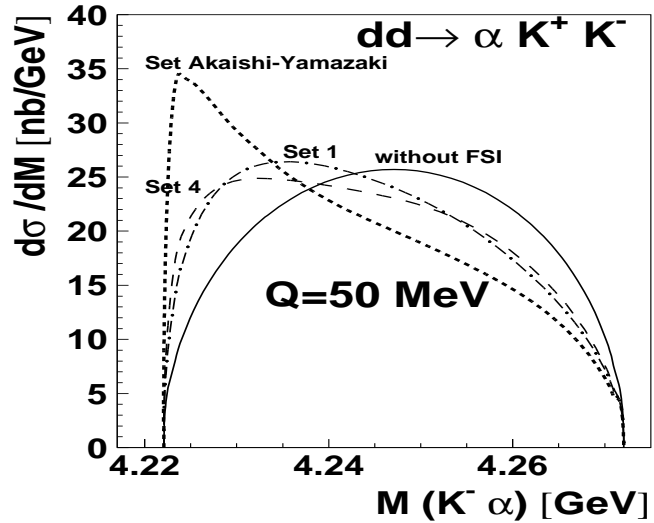,width=8.5cm,height=6.9cm}}
\vspace*{-3mm} \caption{The invariant $K^-\alpha$ mass spectrum
produced in the $dd{\to}\alpha{K^+ K^-}$ reaction at excess energy
50 MeV.  The solid line describes the pure phase space
distribution, while the dash-dotted and dashed lines show our
calculations with $K^-\alpha$ FSI for parameters of Set 1 and 4,
respectively. The short-dashed line shows the result obtained
using the modified $\bar {K} N$ scattering lengths in nuclear
medium.} \label{fig1}
\end{figure}

Akaishi and Yamazaki~\cite{Akaishi} argued that the ${\bar K}N$
interaction is characterized by a strong $I{=}0$ attraction, which
allows the few-body systems to form dense nuclear objects. The optical
potential proposed by Akaishi and Yamazaki for deeply bound nuclear
states contains the following effective $\bar{K} N$ scattering lengths
in the medium: $a^{0}_{\bar{K}N, \, {\mbox{med.}}}=+2.25 +i 0$~fm,
$a^{1}_{\bar{K}N, \,{\mbox{med.}}}=0.48 +i 0.12$~fm. We used these
modified scattering lengths to calculate the enhancement factor for
the $K^- \alpha$ FSI in the $dd \to K^+ K^- \alpha$ reaction. The
short-dashed line in Fig.~\ref{fig1} demonstrates a very pronounced
deformation of the $K^- \alpha$ invariant mass spectrum. Such a strong
in-medium modification of the $\bar{K} N$ scattering length apparently
can be tested at COSY.
}  


\vspace{2ex}                                            
{\small \sl Contact e-mail: kondratyuk@itep.ru}\\
\end{document}